\documentclass[twocolumn,letterpaper,superscriptaddress,showpacs,prl]{revtex4}
\usepackage{graphicx}
\usepackage[usenames]{color}
\usepackage{amsmath}

\begin{document}
\title{Coherent electronic transport through graphene constrictions: \\
sub-wavelength regime and optical analogies}
\author{Pierre \surname{Darancet}}
\affiliation{Institut N\'eel, CNRS \& UJF, Grenoble, France}
\affiliation{European Theoretical Spectroscopy Facility (ETSF), France}
\author{Valerio \surname{Olevano}}
\affiliation{Institut N\'eel, CNRS \& UJF, Grenoble, France}
\affiliation{European Theoretical Spectroscopy Facility (ETSF), France}
\author{Didier \surname{Mayou}}
\affiliation{Institut N\'eel, CNRS \& UJF, Grenoble, France}
\affiliation{European Theoretical Spectroscopy Facility (ETSF), France}

\date{\today}

\begin{abstract}
Graphene two-dimensional nature combined with today lithography
allows to achieve nanoelectronics devices smaller than the Dirac electrons wavelength.
Here we show that in these graphene subwavelength nanodevices the electronic quantum transport properties present deep analogies with classical phenomena of subwavelength optics.
By introducing the concept of {\it electronic diffraction barrier} to represent the effect of constrictions,
we can easily describe the rich transport physics in a wealth of nanodevices:
from Bethe and Kirchhoff {\it diffraction} in graphene slits,
to Fabry-Perot {\it interference} oscillations in nanoribbons. 
The same concept applies to graphene quantum dots and 
gives new insigth into recent experiments \cite{Geim} on these systems.
\end{abstract}

\maketitle

%
Analogies play a preminent role in physics. They allow the transfer of notions and concepts from one field
to another, thus leading to deeper insights and advances in both fields. 
In particular, attempts to explore an analogy between the fields of quantum transport and optics
have been suggested in the past, both in the direction of optics, for instance in order to understand 
coherent multiple scattering of light \cite{Wolf}, and also in the reverse direction \cite{Datta}.
Optical concepts have also been exploited to explain different properties in graphene \cite{Altshuler}. In this work we  develop  the  analogy between quantum transport and optics  beyond those attempts and
bridge coherent electronic transport with sub-wavelength optics.
Indeed graphene \cite{Novoselov,Kim,Berger}, either in the exfoliated or in the epitaxial form, has extraordinary coherence properties.
The electron mean free-path can be of the order of a micron at room temperature
and the Dirac electrons wavelength can be up to 100 nm or even larger \cite{A}.
From a technological point of view, graphene can be manipulated by ordinary lithography
to realize devices as small as few nanometers \cite{B} thus achieving a sub-wavelength regime.
This suggests that an analogy with sub-wavelength optics can reveal fruitful
in order to have a better insight into graphene transport properties.
We believe that this approach could ease the task to find a solution to the formidable
quantum transport problem of calculating the conductance characteristics in nanodevices.

We will show that, by introducing the hybrid ``electronics-optics'' concept of  {\it electronic diffraction barrier}
to represent constrictions at contacts, we can describe in an easy way 
the rich transport physics in a wealth of sub-wavelength graphene nanodevices: 
from the simplest systems such as slits between two half graphene planes, 
to graphene nanoribbons or quantum dots sandwiched in between semi-infinite graphene sheets as leads
\cite{Bethe,Ebbesen,Ozbay,Barnes,Cavity2,Takakura,Suckling}. 
Our methodology relies on an exact numerical calculation of the
conductance within a tight-binding \cite{Wallace} and a Landauer quantum
transport formalisms \cite{Datta}. 
The novelty here introduced is that the Landauer
equation is solved via a new recursive numerical algorithm \cite{MayouRecursion}.

Our results give new insights on the conductance characteristics of
experimentally synthetized graphene quantum dots \cite{Geim,Ensslin} where a chaotic Dirac billiard 
behaviour has been recently observed \cite{Geim}. 
The diffraction barrier concept explains electron localization
and allows us to provide a new interpretation of experimentally
observed features 
observed into this system  
as due to diffraction phenomena occurring at the divergent wavelength 
of electrons at the Dirac point.
We believe that the analogy can be further pushed to play a fruitful and key role in
the design of new graphene nanodevices exhibiting other original and interesting characteristics.
We also point out that the present findings are not only specific of graphene Dirac
electrons but could be extended to other sub-wavelength devices made by 
2D electron gas such as GaAs/AlAs or in molecular electronics.

\begin{figure}
  \includegraphics[clip,width=0.40\textwidth]{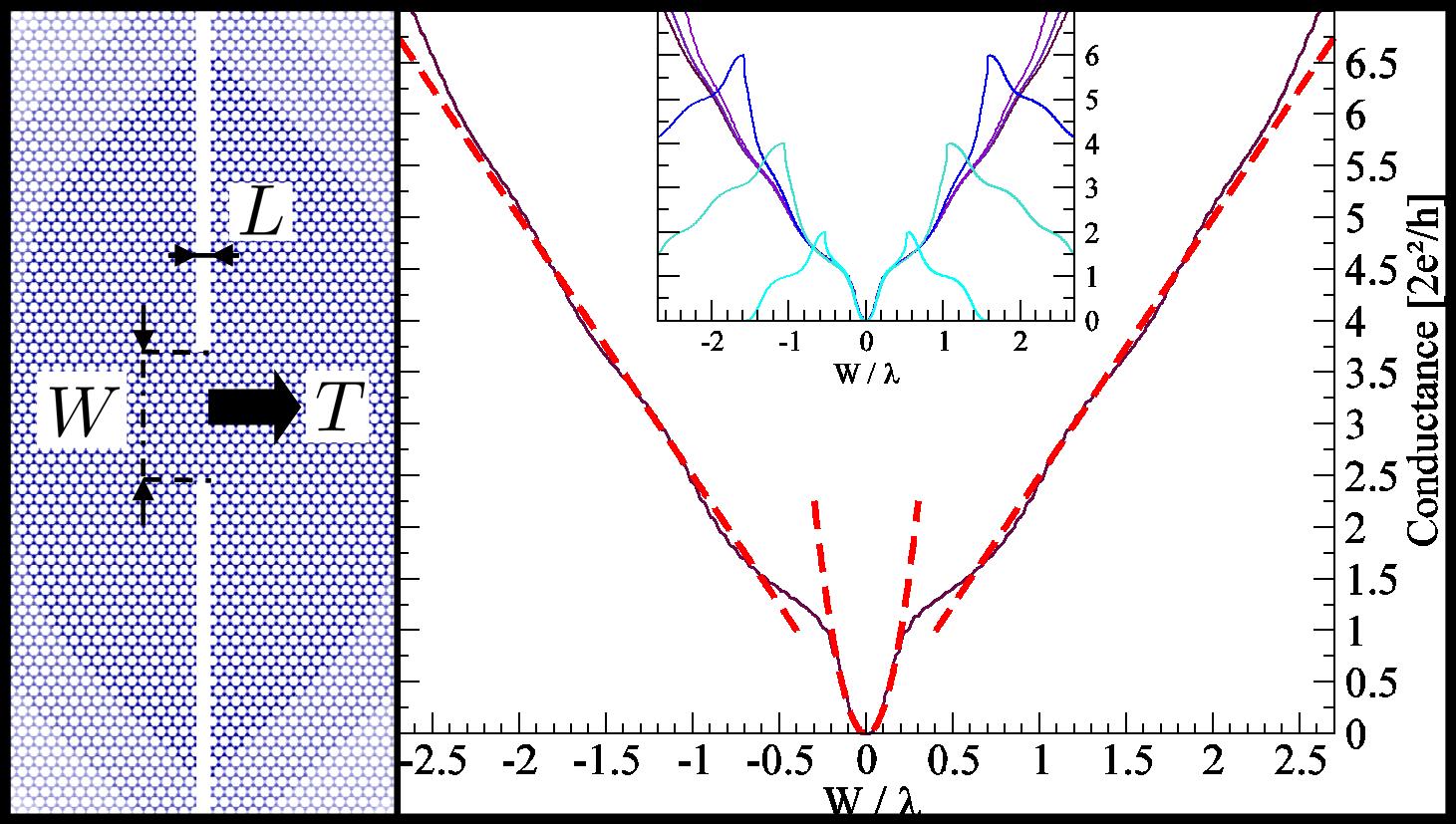}
  \caption{Diffraction of electrons through a graphene slit. 
  Left: schematic view of the geometry of a graphene slit. 
  Right: universal scaling law of the graphene slit conductance as a function of $W/\lambda$. 
  This shows the Bethe-like $\lambda \gg W$ quadratic regime and the
  Kirchhoff-like $\lambda \ll W$ linear regime (see text). The inset
  shows the  conductance 
  for slits of widths $W \simeq 0.8 p$ nm with $p=1,2,3...$}
  \label{diffraction}
\end{figure}

\textit{Conductance of graphene slits and diffraction by an aperture} --
The archetype experiment showing an optical diffraction behaviour
is the transmission through a slit or hole.
We considered slits made into a graphene sheet (Fig.~\ref{diffraction})
\footnote{Here the edges of the graphene half-planes are taken of the armchair type.}
and calculated the conductance. We took into account 
several slits differing by the width $W$, from the thinnest
consisting only of a single \textit{motif}, {\it i.e.} a single hexagon.
We introduce the wavelength $\lambda$ of the incident Dirac
electrons, related to the channel energy $E$ via 
    $\lambda = \frac{2\pi}{k} = \frac{h v_{\rm F}}{E}$
where $k$ is the Dirac electrons wavevector, $E
= \hbar k v_{\rm F}$ is the energy, linearly dispersed in graphene with a
Fermi velocity of $v_{\rm F} \simeq 10^{6}$ ms$^{-1}$
\cite{Kim}.

In Fig.~\ref{diffraction} we report the conductance of the slits
as a function of $W/\lambda$.
We first remark (Fig.~\ref{diffraction} inset) an universal scaling behaviour
independent of the slit width $W$ and depending only on the ratio $W/\lambda$.
This scaling breaks down when the wavelength is $\lambda \lesssim 4$~nm,
that is the energy is $E \gtrsim 1$~eV.
Indeed, as shown below, this scaling is intimately related to the
scaling properties of the Dirac equation which in graphene
is valid only in the low energy or long wavelength limit.

We then notice that in the limit $W/\lambda \ll 1$, the
conductance is quadratic, with the conductance $g(W/\lambda)\simeq 50 (e^
{2}/h)(W/\lambda)^{2}$, while in the opposite limit $W/\lambda \gg 1$, the
conductance turns out to be linear with  $g(W/\lambda)\simeq 5 (e^
{2}/h)(W/\lambda)$. The crossover between the two regimes occurs around
$W/\lambda\simeq 0.2 \sim 0.5$. This promptly reminds of an analogy with
classical optics and hence offers an immediate interpretation: {\it
the conductance response of a graphene slit is a clear manifestation
of a diffraction phenomenon}. Indeed we can identify two different
diffraction regimes: for wavelengths $\lambda \gg W$, much larger
than the aperture, we observe a {\it Bethe-like diffraction regime}
\cite{Bethe} with a slit transmitted energy proportional to the
square of the aperture. On the other hand, for $\lambda \ll W$ one
has a {\it Kirchhoff-like diffraction regime}, the transmitted energy
being proportional to the aperture. Note that weak-localization effects in quantum billards can also reveal diffraction effects in the Kirchoff regime at the lead mouth \cite{Ludger}. In the short wavelength regime the phenomenon of diffraction is a perturbation and a semi-classical description of the electrons as ballistic wavepackets is applicable,
thus leading to a transmitted current proportional to the section of the slit $W$.
By contrast, in the subwavelength regime the transmission is lower than 
given by the semi-classical picture and can be viewed as a tuneling process.

In the ordinary optical diffraction of the slit transmitted intensity,
an universal scaling behaviour is observed due to the scaling invariance of Maxwell
equations. Here it stems from the fundamental scaling invariance of
the graphene 2D Dirac equation. Indeed if a spinor $\psi(r)$ satisfies the Dirac equation for the energy $E$ and wavelength $\lambda$ then  the spinor $\tilde{\psi}(r) = \psi(xr)$ satisfies the Dirac equation for the energy $\tilde{E}=E/x$ and wavelength $x \lambda$. This leads to the fact that
the conductance undergoes a scale invariance $g(x\lambda, xW) =
g(\lambda,W)$. As a consequence the conductance only depends on a reduced argument
$W/\lambda$, $g(\lambda,W)=g(W/\lambda)$.
\footnote{The argument is as follows.
Let us consider a system where each half plane of graphene is replaced
by a ribbon of width $\tilde{W}$. The  symmetry axis of the ribbon is
perpendicular to the slit, passing by the center of the slit. The
conductance for half graphene plane is
$g(\lambda,W)=lim_{\tilde{W}\rightarrow
\infty}g(\lambda,\tilde{W},W)$. We consider a continuous type model of graphene circuit
\cite{Footnote2}. In this model the mass term of the Dirac equation is zero in the graphene zone and infinite outside. This confines the electrons in the graphene zone. Then there is a one to one correspondence between
the scattering states at wavelength  $\lambda$ ribbon width $\tilde{W}$,
slit width $W$  and those at wavelength  $x\lambda$, ribbon width
$x\tilde{W}$  and slit width $xW$. This means that the conductance
satisfies $g(\lambda,\tilde{W},W)=g(x\lambda,x\tilde{W},xW)$ and as a
consequence $g(x\lambda, xW) = g(\lambda,W)$}. Let us note that a  scaling relation can also exist for other geometries. In that case the conductance of the circuit will depend only on the ratio between the characteristic lengthes of the circuit and the wavelength $\lambda$.

\begin{figure}
  \includegraphics[clip,width=0.40\textwidth]{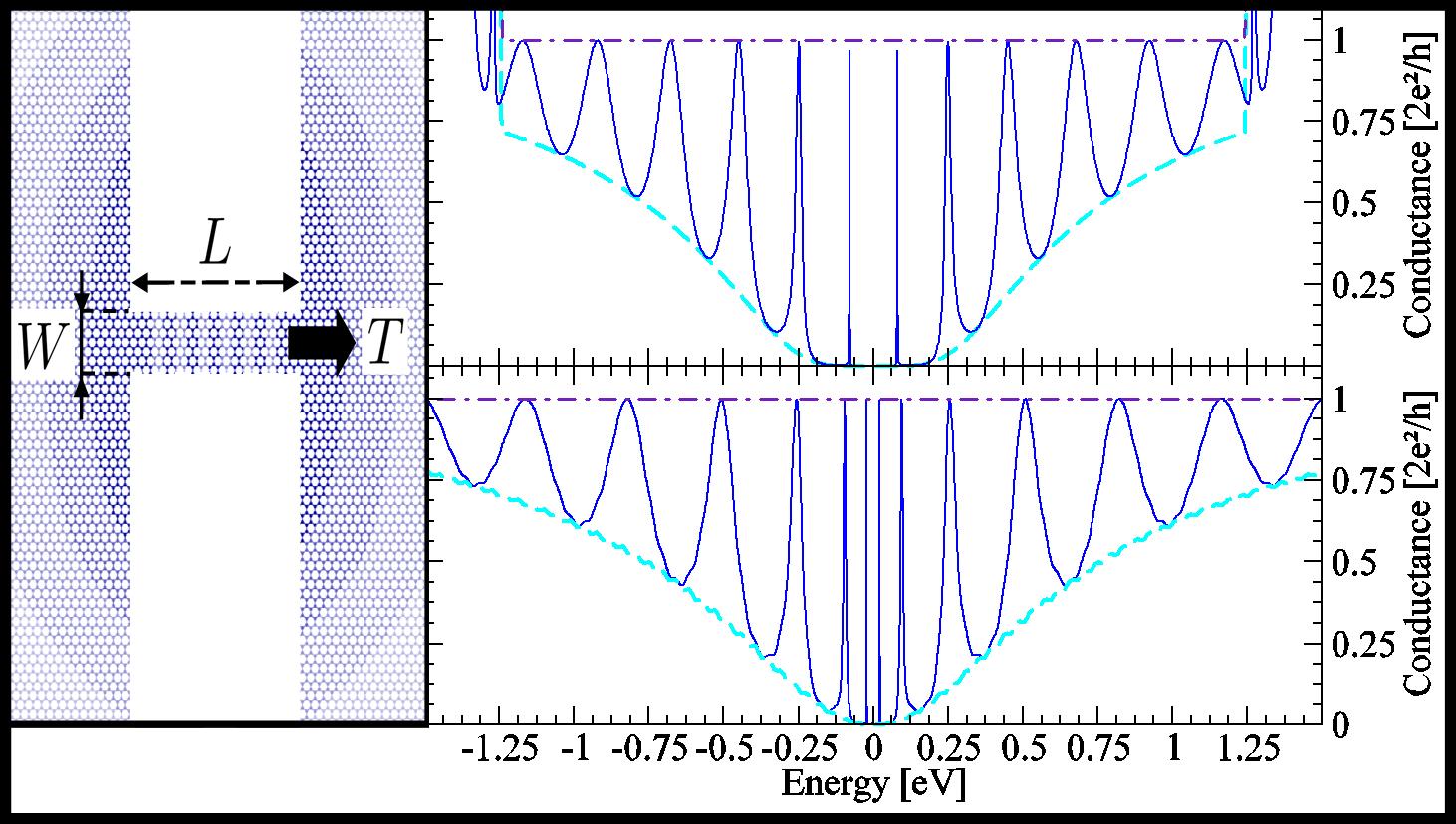}
  \caption{Graphene nanoribbons as Fabry-Perot interferometers.
  Left: schematic view of nanoribbons contacted to graphene half-planes. 
  Right: Fabry-Perot oscillations inf the conductance of armchair (upper panel) 
  and zigzag (lower panel) nanoribbons.
  The dashed line which almost perfectly envelops along the oscillations minima, is the plot
  of Eq.~(\ref{minimaenvelope}) with $R(E)$ taken from Fig.~\ref{junctions}}
     \label{nanoribbons}
\end{figure}

\begin{figure}
  \includegraphics[clip,width=0.40\textwidth]{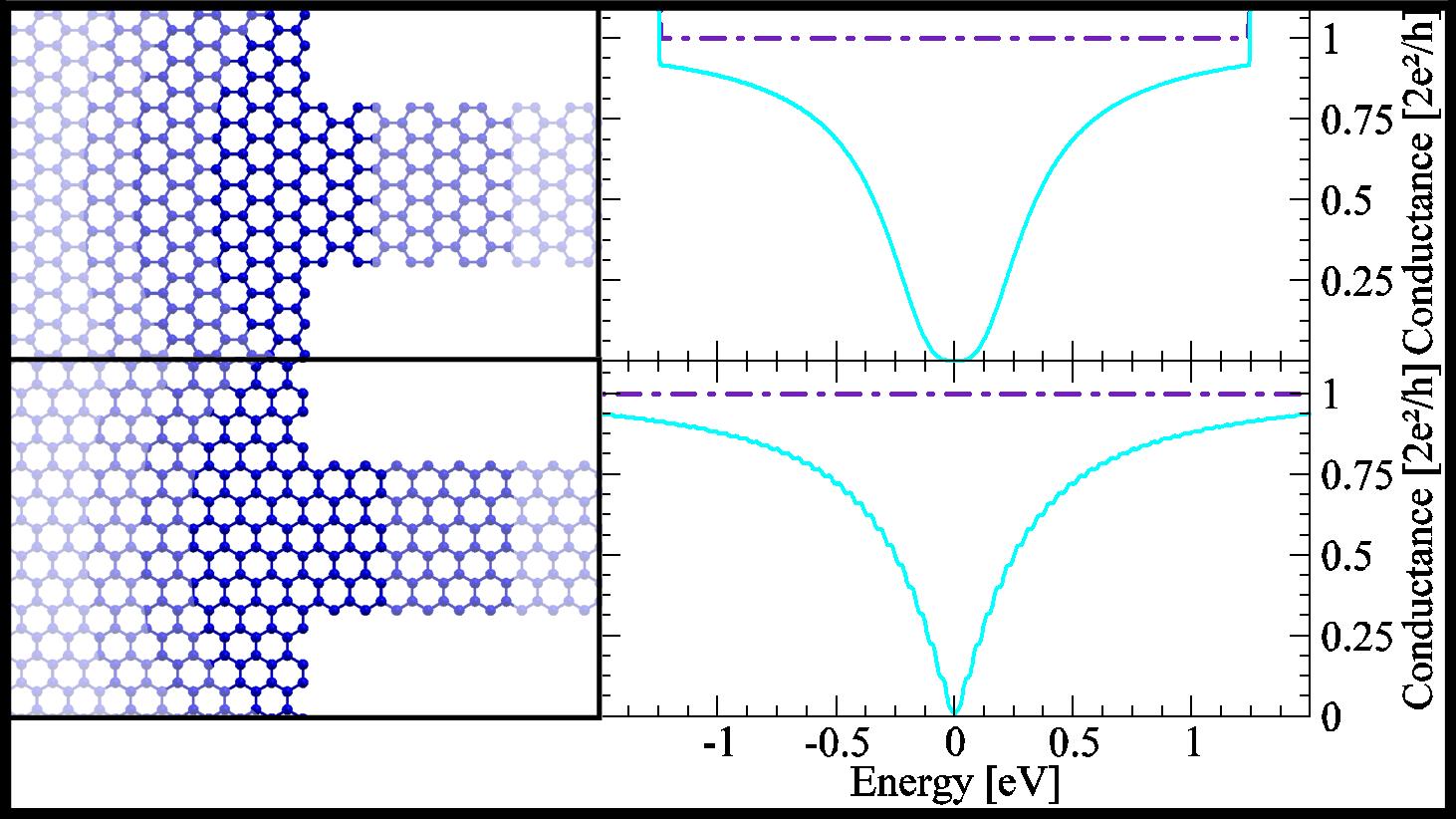}
  \caption{\textit{Electronic diffraction barrier} at a contact. 
  Left: the contact geometries of the armchair (upper panel)  
  and zigzag (lower panel) semi-infinite ribbons coupled to graphene half-planes. 
  Right: Conductance $g$ of the device represented on the corresponding left side. 
  $g=T 2e^{2}/h$ with $T$ the transmittance of the diffraction barrier.
}
     \label{junctions}
\end{figure}

\textit{Nanoribbons as Fabry-Perot interferometers and subwavelength waveguides} -- Here we study the quantum transport response of finite-length metallic graphene nanoribbons
and show that they present an oscillating response, typical of optical interferometers,
and behave as subwavelength waveguides.
We consider two ribbons, a zigzag and an armchair
both chosen with a metallic character and hence an available
conductance channel at the Dirac energy \cite{dresselhaus}. The geometry is presented  in Fig.~\ref{nanoribbons} and we have taken a length of $6$ nm for the armchair nanoribbon
and $3$ nm for the zigzag (see Fig.~\ref{junctions} for the detailed geometry of each junction). The calculated conductance for both
ribbons is shown in Fig.~\ref{nanoribbons} where we again observe a conductance going to 0 at the Dirac point.
However in this case, the most evident features are
large amplitude {\it oscillations} in the conductance, from maxima values of 1
to minima disposed along an envelope function.
These systems behave as  Fabry-Perot cavities, exactly as in subwavelength
optical metallic waveguides \cite{Cavity2,Takakura,Suckling}. To demonstrate this
we show that the nanoribbons calculated oscillating
conductance distributes like the Fabry-Perot standard transmittance
which is the Airy function:

\begin{equation}
     T_{\rm FP}(E) = \frac{1}{1 + F(E) \sin^{2}(\phi(E)/2)}
    \label{T}
\end{equation}
where $\phi(E)= 2 k(E)L + 2 \tilde{\phi}(E)$ is the phase difference
after one loop in the nanoribbon. $k(E)$ is the wavevector of the
Bloch state in the infinite ribbon with energy $E$. $\tilde{\phi}(E)$
is the phase factor acquired at each reflection. $L$ is the length of
the Fabry-Perot interferometer.  $F(E) = 4R(E)/(1-R(E))^{2}$ is the
{\it finesse} coefficient and $R(E) = 1 - T(E)$ is the reflection coefficient
at each end of the nanoribbon, that is at the junction.

The Airy function presents maxima $T_{\rm FP}^{max} = 1$ equal to 1
when $\phi/2$ is an integer $m$ multiple of $\pi$.  For sufficiently
large $L$ the phase $\phi(E)= 2 k(E) L + 2 \tilde{\phi}(E)$ varies
rapidly with energy, as compared to $F(E)$; and the minima occur
essentially when $\sin2(\phi/2)$ is maximum, that is at $\phi/2=m\pi +
\pi/2$, and envelope along the function
\begin{equation}
     T_{\rm FP}^{min}(E) = \frac{1}{1+F(E)} = \frac{(1-R(E))^{2}}{(1+R(E))^{2}}.
     \label{minimaenvelope}
\end{equation}
which is independent of the length of the ribbon. We indeed checked that the envelope functions of
the conductance minima, as expected from
Eq.~(\ref{minimaenvelope}), is independent of the ribbon length. In particular the minimum of 
the transmission tends to zero at the Dirac energy, which indicates that $R(E)$ tends to $1$  at zero energy.

Another feature of the Fabry-Perot resonnances concerns their width. The peaks full width at half maximum is $\delta E_{\rm FWHM} = \delta \phi \frac{dE}{d\phi} = \delta \phi  \frac{v(E)}{2L}$ where $\delta \phi =
2(1-R)/\sqrt{R}$ and $v(E)$  is the group velocity at energy $E$. The velocity is bounded,
such as the peaks look thin when $R$ is close to 1, and broaden when $R \rightarrow 0$. In our case when $E\rightarrow 0$
the peaks seem to have negligible width, indicating again  that
$R\rightarrow 1$, and broaden for the higher energies, indicating
lower values of $R$.

The relative variation of the conductance ({\it i.e.}
the ratio between the maximum and minimum conductance between two
peaks)  is also maximum close to the Dirac energy, due to the nearly
perfect confinement at this energy ($R(E)\simeq 1$). We consider in the following  that resonances are well defined if the maximum  to minimum ratio is greater than $2$. 
In our calculations this corresponds to a criterion $\lambda/W  \gtrsim 3\sim 5$ 
where $W$ is the width of the nanoribbon.

\textit{The concept of electronic diffraction barrier at a constriction} --
Fabry-Perot oscillations occur due to the reflection at the junctions/ends
of the nanoribbons. We now analyse this reflection and the response of the elementary junction.
We consider the same metallic zigzag and armchair ribbons
as before, but take them as semi-infinite (Fig.~\ref{junctions}).  If the system
were ballistic (infinite ribbon, no contact resistance and no reflexion),
we would expect to observe a conductance equal to 1 quantum of conductance
($2e2 /h$) (dot-dashed line in Fig.~\ref{junctions}) in the range of
the Dirac energy. Instead, the calculated transmittance $T(E)$
through the junction (continous line in Fig.~\ref{junctions}) is
observed to drop to zero at the Dirac energy, no matter the
chirality, exactly like in slits and the nanoribbons. The only effect of the ribbon electronic structure is
to produce a little difference in the characteristics, with the
zigzag conductance showing a more cusp-like feature at the Dirac
energy.

Since the associated resistance
is independent of the ribbon length, this behaviour is a genuine
manifestation of the contact resistance at the junction
\footnote{We are considering metallic nanoribbons. For non-metallic nanoribbons
the resistance is  exponentially increasing with length.}.
The junction conductance we have here calculated represents the characteristics
of the elementary {\it electronic diffraction barrier} which is at the basis of the
universal behaviour observed in all the devices hereby studied,
slits, nanoribbons or quantum dots. 
Indeed coming back to the nanoribbon case, if in Fig.~\ref{nanoribbons} we plot $(1-R(E))^{2}/(1+R(E))^{2}$ (dashed curve) with $R(E) = 1 - T(E)$ as extracted from Fig.~\ref{junctions} for the elementary junction,
we find that the curve turns out to be the envelope of the Fabry-Perot
oscillations minima in the nanoribbon conductance, as expected from Eq.~(\ref{minimaenvelope}).
This exemplifies the concept of $\textit{electronic diffraction barrier}$ at a constriction and 
gives a quantitative example of such a barrier.

\begin{figure}
     \includegraphics[clip,width=0.40\textwidth]{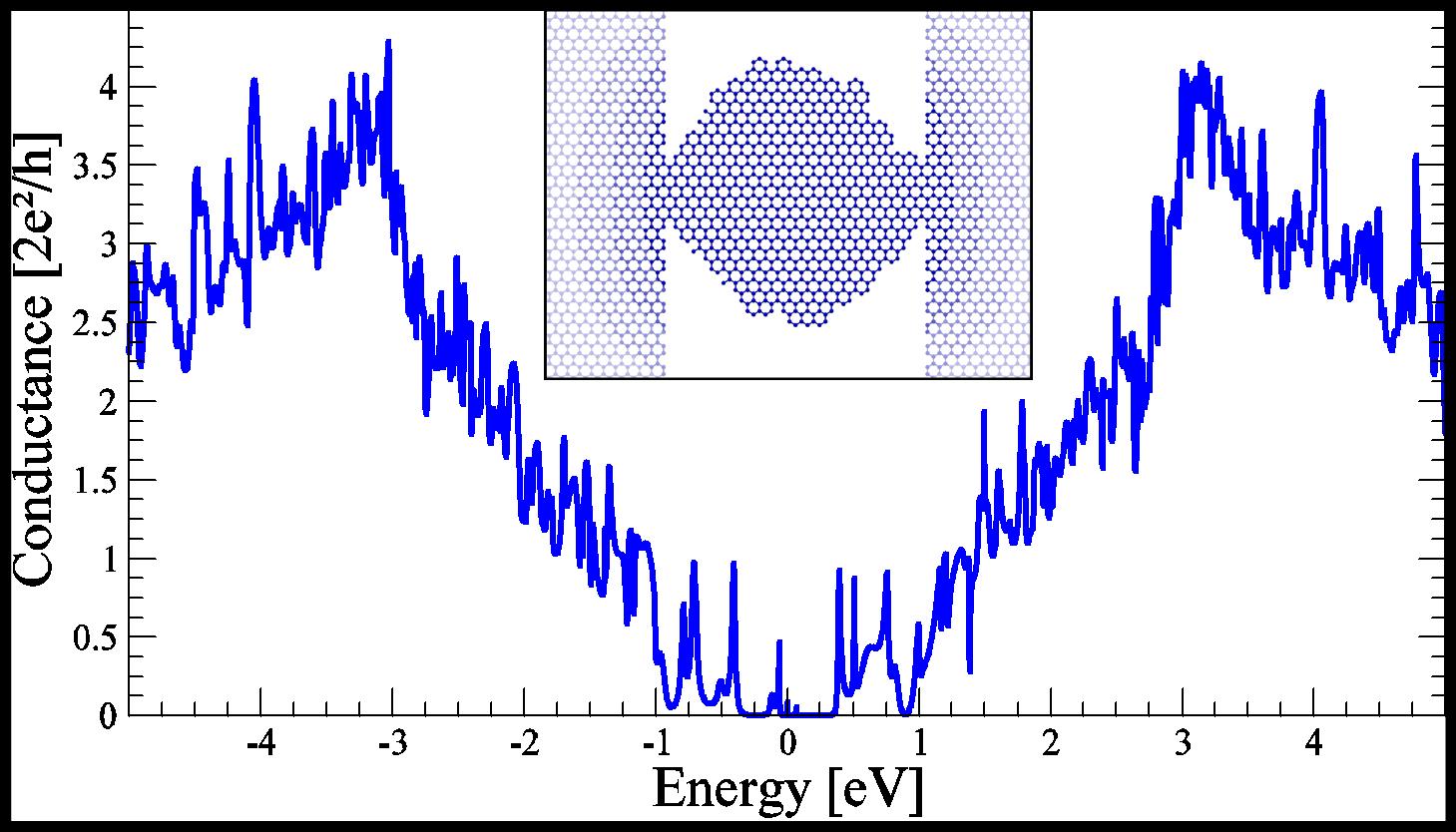}
     \caption{Conductance of the quantum dot, showing well defined
resonances close to the Dirac energy.  Inset: geometry of the
irregular shaped quantum dot coupled to the graphene half-planes. }
     \label{quantumdot1}
\end{figure}

\textit{Conductance of a quantum dot} --
We now turn to more complicated structures and
consider a standard nanoelectronics quantum dot
(Fig.~\ref{quantumdot1}) consisting of a purposely
irregular shape graphene nanostructure contacted via small apertures
to the two half-planes graphene leads.
Applying again the concept of electronic diffraction barrier,
we expect that the quantum dot is weakly coupled to the leads when the
energy of the electrons in the graphene sheet is close to the Dirac
energy, {\it i.e.} when their wavelength is sufficiently large.
We note that  the conductance of this system (Fig.~\ref{quantumdot1})
presents a maximum around $3$ eV  which is  the energy of the
Van-Hove singularity of bulk graphene where the density of states
diverges. We also note many irregular peaks superimposed to the
envelope, clearly showing characteristic resonances of the quantum
dot to be associated to the particular (irregular) shape. The
conductance decreases when going to the Dirac point, which is a
signature of the lowest number of available states and
of a diffraction barrier at the contact constrictions.
The maximum to minimum transmission ratio between adjacent peaks is
much larger in the vicinity of the Dirac energy than at other
energies. This is also observed in the nanoribbons studied previously.
Again this can be explained by the presence of two diffraction
barriers at the contacts
which are nearly completely reflecting ($R=1$) at the Dirac energy. This leads
to well defined states within the quantum dot, and thus well defined
conduction channels through these states. 
If we take as criterion to identify a range with well defined resonances
the same we already introduced (maximum to minimum ratio greater than $2$), 
then we find that the critical wavelength is of order of $\lambda \simeq 4$nm  for a width $W = 1.2$nm. This is the same range of values of $\lambda/W$ as in nanoribbons, \textit{i.e.} $\lambda/W  \gtrsim 3\sim 5$.

\textit{Experimental implications} -- Our quantum dot computed characteristics (Fig.~\ref{quantumdot1}) looks
very similar to the experimental result of Ponomarenko et al. (Ref.~\cite{Geim}, Fig.~1). In those experiments the relation between voltages $V$  (here $V$ is the voltage variation with respect to the Dirac point voltage expressed in Volts)  and electronic wavelength $ \lambda$ is  $\lambda\simeq 130nm/\sqrt{V}$  \cite{NovoselovPrivate}. From that  one deduces that in their experiment the resonances are well  defined for a wavelength $\lambda$ to aperture $W$  ratio $\lambda/W  \gtrsim 3\sim 5$ in reasonable agreement with the numerical results presented here.  They reported also the low bias conductance of constrictions of 20 nm (Ref.~\cite{Geim}, Fig.~S2),  similar to the slits studied here. First of all, after our study of slits the Kirchoff-like (linear) regime is reached at voltages $V \geq 10$Volts and for higher voltages we predict for this $20 nm$ slit  a conductance $g(W/\lambda)\simeq 5 (e^
{2}/h)(W/\lambda)\simeq 0.75 (e^{2}/h) \sqrt{V}$.  This is consistent with the experimental results and for instance
at 100 Volts from the voltage corresponding to the Dirac point they measure $G \simeq  5.5 e^2/h$
(Ref.~\cite{Geim}, Fig.~S2),
while
we  predict $G \simeq 7.5 e^2/h$, in good agreement with the experiment.
On the other hand, the comparison of our results
with their smaller $10$ nm constriction is not such good.
Other mechanisms should influence the electronic transport in this case.
This is in agreement with the evidence of thermally activated
transport for the smaller experimental constriction.

%

To conclude, our work establishes an important link between
nanoelectronics and subwavelength optics which could be tested
in today's technology graphene nanodevices, thank to its remarkable properties.
The concept of \textit{electronic diffraction barrier}
is central to the understanding of transport properties in
the subwavelength regime. It allowed us to get insights
into the properties of graphene junctions, slits, nanoribbons, and
quantum dots and explain their conductance characteristics
in terms of diffraction and interference phenomena.
This analogy can be further pushed forward in the design
of devices with new properties.

\textit{Acknowledgements} -
This work has benefited from exchanges with many colleagues. We thank
in particular  C.~Berger, X. Blase, F.~Balestro, D.~Feinberg, 
L.~L{\'e}vy, L.~Magaud, P.~Mallet,  C.~Naud, T.~Lopez-Rios,  P.~Qu{\'e}merais,
F.~Varchon, J.~Y.~Veuillen, W.~Wernsdorfer, L. ~Wirtz. Computer time has been granted by
IDRIS and Ciment/Phynum.


\begin{thebibliography}{99}

\bibitem{Geim}
L.~A. Ponomarenko et al., Science {\bf 320}, 356 (2008).

\bibitem{Wolf}
P.-E. Wolf and G.  Maret,
Phys. Rev. Lett. {\bf 55}, 2696 (1985).

\bibitem{Datta}
S. Datta, 
{\it Electronic Transport in Mesoscopic Systems}, Cambridge University Press, Cambridge 1995.


\bibitem{Altshuler}
V.~V. Cheianov, V. Fal'ko and B.~L. Altshuler,
Science \textbf{315}, 1252 (2007).


\bibitem{Novoselov}
K.~S.~Novoselov et al., Nature {\bf 438}, 197 (2005).

\bibitem{Kim}
Y.~Zhang, Y.-W.~Tan, H.~L.~Stormer and P. Kim, Nature {\bf 438}, 201

\bibitem{Berger}
C.~Berger et al., Science {\bf 312}, 1191 (2006).


\bibitem{A}
S.~V. Morozov et al., Phys. Rev. Lett. {\bf 100}, 016602 (2008).

\bibitem{B}
B. \"Ozyilmaz et al., Phys. Rev. Lett. {\bf 99}, 166804 (2007).



\bibitem{Bethe}
H.~A. Bethe, Phys. Rev. {\bf 66}, 163 (1944).


\bibitem{Ebbesen}
T.~W. Ebbesen et al. Nature {\bf 391}, 667 (1998).

\bibitem{Ozbay}
E. Ozbay,
Science {\bf 311}, 189 (2006).

\bibitem{Barnes}
W.~L. Barnes, A. Dereux and T.~W. Ebbesen,
Nature {\bf 424}, 824 (2003).

\bibitem{Cavity2}
P. Qu\'emerais, A. Barbara, J. Le~Perchec and T. L\'opez-Rios,
J. App. Phys. \textbf{97}, 053507 (2005).

\bibitem{Takakura}
Y. Takakura, 
Phys.~Rev.~Lett. {\bf 86}, 5601 (2001).

\bibitem{Suckling}
J.~R Suckling et al.
Phys.~Rev.~Lett. {\bf 92}, 147401 (2004).


\bibitem{Wallace}
P.~R. Wallace, Phys. Rev. {\bf 71}, 622 (1947).

\bibitem{MayouRecursion}
P. Darancet, V. Olevano and D. Mayou, to be published.
P. Darancet, PhD thesis.


\bibitem{Ensslin}
C. Stampfer et al., Appl. Phys. Lett. {\bf 92}, 012102 (2008).

\bibitem{Ludger}
I.Brezinova, C.Stampfer, L. Wirtz, S. Rotter, J. Burgd \"orfer,  Phys.Rev.B {\bf 77}, 165321 (2008).

\bibitem{Footnote2}
N.~M.~R. Peres, A.~H. Castro Neto and F. Guinea,
Phys.~Rev.~B {\bf 73}, 241403(R) (2006).

\bibitem{dresselhaus}
K. Nakada, M. Fujita, G. Dresselhaus and M.~S. Dresselhaus, 
Phys.~Rev.~B {\bf 54}, 17954 (1996).










\bibitem{NovoselovPrivate}
K.~S. Novoselov, private communication.







\end{thebibliography}
\end{document}